\documentclass[10pt,aps,prd,superscriptaddress,nofootinbib,nobibnotes,
    longbibliography,floatfix,twocolumn]{revtex4-2}

\usepackage{bm}
\usepackage{mathtools,
amsmath,
amssymb,
amsfonts,
mathrsfs,
chngcntr,
multirow}

\let\cc\corresponds
\let\corresponds\relax
\usepackage{mathabx}
\let\corresponds\cc

\usepackage{journals}
\usepackage{siunitx}
\usepackage[utf8]{inputenc}
\usepackage[T1]{fontenc}
\usepackage[dvipsnames]{xcolor}
\usepackage[unicode]{hyperref}
\hypersetup{colorlinks=true, citecolor=MidnightBlue,
            linkcolor=MidnightBlue, urlcolor=MidnightBlue, linktocpage=true}
\usepackage[capitalize]{cleveref}

\newcommand{\orcid}[1]{\href{https://orcid.org/#1}{\includegraphics[width=10pt]{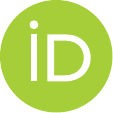}}}

\begin{document}

\title{Combining simulation-based inference and universal relations\\  for precise and accurate neutron star science}

\author{Christian J.\,Kr\"uger \orcid{0000-0003-2672-2055}}
\email{chr.krueger@uni-tuebingen.de}
\affiliation{Theoretical Astrophysics, IAAT, University of T\"ubingen, 72076 T\"ubingen, Germany}

\author{Sebastian H.\,V\"olkel\,\orcid{0000-0002-9432-7690}}
\email{sebastian.voelkel@aei.mpg.de}
\affiliation{Max Planck Institute for Gravitational Physics (Albert Einstein Institute),
D-14476 Potsdam, Germany}

\date{\today}

\begin{abstract}
In this work, we propose a novel approach for identifying, constructing, and validating precise and accurate universal relations for neutron star bulk quantities.
A central element is simulation-based inference (SBI), which we adopt to treat uncertainties due to the unknown nuclear equation of state (EOS) as intrinsic non-trivial noise.
By assembling a large set of bulk properties of non-rotating neutron stars across multiple state-of-the-art EOS models, we are able to systematically explore universal relations in high-dimensional parameter spaces.
Our framework further identifies the most promising parameter combinations, enabling a more focused and traditional construction of explicit universal relations.
At the same time, SBI does not rely on explicit relations; instead, it directly provides predictive distributions together with a quantitative measure of systematic uncertainties, which are not captured by conventional approaches.
As an example, we report a new universal relation that allows us to obtain the radius as a function of mass, fundamental mode, and one pressure mode. 
Our analysis shows that SBI can surpass the predictive power of this universal relation while also mitigating systematic errors. 
Finally, we demonstrate how universal relations can be further calibrated to mitigate systematic errors accurately. 
\end{abstract}

\maketitle

\section{Introduction}


Modeling neutron stars comes with significant challenges due to uncertainties in the underlying nuclear equation of state (EOS) at high densities. 
In the simplest, non-rotating case, a given EOS specifies one family of neutron star models, e.g., as a function of the central density. 
One well-established approach that bypasses EOS uncertainties is the construction of universal relations. 
They establish empirical, EOS-insensitive relations between neutron star bulk properties such as mass, radius, and oscillation modes, amongst others. 
From first principles, it is not obvious in the majority of cases why the complexity of general relativity, especially its dynamics, and the variety of different EOS allow for such simple relations. 
Pioneering works for isolated neutron stars can be found in Refs.~\cite{Andersson:1997rn, Benhar:1998au, Gaertig:2008uz, Breu:2016ufb, Musolino:2023edi}. 
In recent years, universal relations have also been developed to study binary neutron star mergers, for example, to estimate the maximum mass~\cite{Rezzolla:2017aly}, the threshold mass~\cite{Bauswein:2013jpa}, the disc mass~\cite{Kruger:2020gig}, or relations involving the peak frequency~\cite{Vretinaris:2019spn}. 

Despite their advantages, universal relations also come with challenges and limitations. 
Three of the most relevant ones, especially in the context of our work, are the following. 
First, constructing a new functional form that connects (relevant) bulk properties can be tedious and may require vast experience with existing universal relations. 
Second, even if a robust relation is found, the systematic errors are not reliably accounted for by the uncertainty in the fit parameters of the universal relations, which we outline further in Sec.~\ref{methods_ur}. 
Third, the general robustness of the relation also depends on how thoroughly EOS uncertainties and biases are accounted for, e.g., through the size of the viable EOS set or the number of agnostic parametrizations considered. 
Along the line of automating the search for promising parameter combinations, recent progress has been made by investigating data analysis methods for non-rotating stars and a small sample of EOS in Ref.~\cite{Manoharan:2023atz}. 
Universal Relations for rapidly rotating neutron stars using supervised machine-learning techniques have been reported in Refs.~\cite{Papigkiotis:2023onn, Papigkiotis:2025cjy, Papigkiotis:2025dka}. 
One application to mitigate systematic errors for simulated measurements for future gravitational wave measurements related to the tidal deformability was reported in Ref.~\cite{Kashyap:2022wzr}. 

In this work, we propose a new strategy centered on simulation-based inference (SBI) that, in principle, overcomes all traditional limitations. 
It is a modern and powerful tool for data-analysis problems, see Ref.~\cite{doi:10.1073/pnas.1912789117} for a recent review and Refs.~\cite{Dax:2021tsq,Dax:2022pxd,Dax:2024mcn,Crisostomi:2023tle} for applications to gravitational waves. 
SBI can have significant advantages compared to traditional Bayesian methods~\cite{sivia2006data}, because they rely on specifying the likelihood function, which may, in general, not be known or must be approximated. 
SBI circumvents this limitation and, thus, is sometimes also called likelihood-free inference. 
Instead of first specifying an explicit likelihood model, SBI only requires simulations or observations expressed directly as functions of the underlying parameters to be inferred.
Since it is often easier to compute an underlying model and then add complicated noise realizations, SBI can be used in more complex scenarios. 
To reduce biases due to a limited set of implemented EOSs, we produce data based on many realizations of multiple state-of-the-art models, allowing for generous coverage~\cite{Read:2008iy, Greif:2018njt, OBoyle:2020qvf, Annala:2019puf, Hinderer:2007mb}.

The key idea of our approach for using SBI for neutron star universal relations is to interpret the systematic errors, due to the variations within the underlying broad set of EOSs and the finite accuracy of universal relations, as intrinsic, non-trivial noise, i.e., as an aleatoric uncertainty.
Since it is related to the uncertainties in the EOS, we will henceforth refer to it as ``EOS noise''.
Sampling the ``posterior'' distribution for a given set of input parameters not only provides an estimate of the most likely output given by its mean value (corresponding to a traditional universal relation), but also quantifies systematic uncertainties, i.e., the EOS noise. 
Having control over systematic errors is crucial for dealing with biases and quantifying the reliability of the relations. 

We find that SBI is valuable and easy to incorporate into strategies for identifying promising parameter combinations in a first step. 
In a second step, promising parameter combinations can then be used to start a more targeted and conventional search for new universal relations. 
During our search, we detect the presence of several known universal relations and also report a new universal relation for the radius $R = R (M, f, p_1)$, which is well beyond percent accuracy. 
Comparing it with the SBI predictions, we find that SBI can outperform the universal relation with sufficient training data, and provides radius estimates of only a few tens of meters. 
At the same time, we also report that the SBI estimates for the systematic errors of the predictions are reliable. 
Finally, we show that the standard way of constructing universal relations does not provide useful estimates for systematic errors, while a newly proposed calibration procedure does. 
Here, we introduce an effective EOS noise error for the radius leading to accurate estimates.

Unless noted otherwise, we employ units in which $c = G = M_\odot = 1$.

\section{Methods}

\subsection{Generation of neutron star data}
\label{ssec:data-set}

We aim to discover robust universal relations between neutron star bulk quantities, including the $p_1$-mode frequency, which has not often been considered in this context yet. In particular, our list of bulk quantities consists of the gravitational mass $M$, radius $R$, moment of inertia $I$, tidal deformability $\Lambda$, and the frequencies $f$ and $p_1$ of the $f$(undamental)-mode and the first $p$(ressure)-mode, respectively. We restrict ourselves to non-rotating neutron stars, since the $p$-mode frequencies of rotating neutron stars are only very difficult to access in sufficient numbers, and the tidal deformability is hitherto unknown; further, we consider only quadrupolar modes, i.e. $l=2$. To reveal relations that are independent of the EOS, we generate random EOS realizations based on four different parametrizations \cite{Read:2008iy, OBoyle:2020qvf, Greif:2018njt, Annala:2019puf}; in fact, we take these EOS parametrizations from a prior study \cite{Kruger:2025sda}.\footnote{All employed EOS parametrizations have been implemented in a C library, using numerical routines from the GNU Scientific Library (GSL) \cite{gsl}.} In total, we consider 1491 different EOS realizations that fulfill basic astrophysical constraints: we demand that the EOS remains causal (i.e., $0 \le c_s^2 < 1$) up to the maximum mass model, which should have a mass of at least $M_\textrm{TOV} = 1.97\,M_\odot$ \cite{Antoniadis:2013pzd};\footnote{The observation of PSR J0740+6620 \cite{Fonseca:2021wxt, NANOGrav:2019jur} suggests a slightly stronger bound $M_{\rm TOV} \ge 2.01\,M_\odot$ when uncertainties are taken into account. The difference to our adopted constraint is small and would not materially affect our results.} the radius $R_{1.6}$ of a $1.6\,M_\odot$ neutron star must exceed $10.6\,\textrm{km}$ \cite{Bauswein:2017vtn}, and the radius $R_{1.4}$ of a $1.4\,M_\odot$ star has to lie within the range $11.5\,\textrm{km} \le R_{1.4} \le 13.5\,\textrm{km}$ \cite{Raaijmakers:2021uju}; finally, the tidal deformability $\Lambda_{1.4}$ of a $1.4\,M_\odot$ neutron star must fall within the interval $120 \le \Lambda_{1.4} \le 800$ \cite{Annala:2017llu, LIGOScientific:2018cki}. For each EOS, we randomly select five neutron star models with masses of at least $1\,M_\odot$ by drawing uniformly in central energy density. This results in a mass distribution with more weight toward higher-mass stars, while still retaining sufficient coverage of low-mass models. We then compute the quantities mentioned above by means of the TOV equations, Hartle's equation \cite{Hartle:1967he}, the Love-number equation \cite{Hinderer:2007mb, Hinderer:2007mberr}, and the standard eigenvalue formulation for mode calculations \cite{Lindblom:1983ps, Detweiler:1985zz, Andersson:1995wu} to an accuracy of at least $10^{-5}$. After discarding models that do not satisfy our mode-identification criteria, the final data set contains 7346 neutron star models.

\subsection{Simulation-based inference}

In our work, we  utilize the popular \texttt{Python} package \texttt{sbi}~\cite{tejero-cantero2020sbi, tejero_cantero_2022_8192532, BoeltsDeistler_sbi_2025}. 
It provides a solid code infrastructure to tackle a variety of SBI problems by using machine-learning techniques such as neural posterior estimation (NPE), neural likelihood estimation, and neural ratio estimation. 
After training on a large set of simulated observations labeled by their underlying parameters, NPE~\cite{Papamakarios:2016ctj,lueckmann2017flexiblestatisticalinferencemechanistic,greenberg2019automaticposteriortransformationlikelihoodfree,deistler2022truncatedproposalsscalablehasslefree} enables direct and fast sampling of an approximate posterior distribution for given input data due to the use of normalizing flows. 
In our work, we adopt NPE and simply refer to a large number of samples as the posterior distribution; when we say SBI in our applications, we refer to the specific implementation through NPE. 

We split our neutron star data described in Sec.~\ref{ssec:data-set} into three sets with sizes of 72\%/8\%/20\% for training, validation, and testing, respectively. We then systematically explore all 602 possible configurations obtained by assigning each of the six neutron star quantities ($M$, $R$, $I$, $\Lambda$, $f$, $p_1$) either to the ``data'' subset, the ``observations'' subset, or to neither. For each case, we initialize the training for NPE. Each case takes only a couple of minutes of training on a standard workstation.
To easily filter for parameter combinations that result in narrow posteriors located very close to the true value, we employ two basic metrics. First, we compute the deviation of the posterior mean from the true value, and secondly, the width of the $68\,\%$ highest-density interval (HDI).
Note that this initial analysis is not sufficient for any proper validation of the NPE; we discuss this in Sec.~\ref{sec_app_results}. 

As a first demonstration of our procedure, we find that the well-known $f$-$I$ \cite{Lau:2009bu}, $I$-Love \cite{Yagi:2013bca} and $f$-Love \cite{Chan:2014kua} relations are accurately reproduced by our trained NPE in the sense that the corresponding combinations of data and observations result in sharp posteriors at the appropriate locations. One of the first published asteroseismology relations (the predecessor to the $f$-$I$ relation) estimating the $f$-mode frequency as $f = f(M, R)$ \cite{Andersson:1997rn} also ranks highly in our analysis; however, we find that the posteriors are somewhat broader and sometimes slightly offset. This finding not only aligns with the somewhat lower accuracy of that relation, but also suggests that the observables $M$ and $R$ cannot be combined in a different way to produce a significantly better estimate for the $f$-mode frequency. 
One novel relation that ranks highly, which we will discuss in more detail in the following, is the dependency $R=R(M, f, p_1)$.

\subsection{Universal relation}
\label{methods_ur}

With the ``educated guess'' that a universal relation $R=R(M, f, p_1)$ should exist, we empirically tried different functional forms to construct it traditionally. 
Due to the higher-dimensional parameter space, finding such a relation, in practice, is not trivial and requires trial and error. 
Using the same data used for SBI training, we report that the relation
\begin{align}
    R
    & = a_0 + a_1 \cdot\sqrt{M} + a_2 \cdot \frac{1}{f} + a_3 \cdot \left( M f \right)^2 + a_4 \cdot \frac{f}{p_1} \,,
    \label{eq:RofMfp1}
\end{align}
can predict the radius with high accuracy. The bulk quantities in this relation are expressed in geometric units, with conversion factors  $G M_\odot / c^2 \approx 1.477\,\text{km}$ for radii and $(G M_\odot / c^3)^{-1} \approx 203.0\,\text{kHz}$ for frequencies. Using a standard least-squares fit employing the Levenberg-Marquardt algorithm, we obtain the numerical values of the best-fit values of $a_i$, as well as the covariance matrix $\sigma$, which encodes their errors. The best-fit coefficients given our data set are $a_0 = -3.312$, $a_1 = 4.864$, $a_2 = 4.360 \times 10^{-2}$, $a_3 = -2.828 \times 10^3$, $a_4 = 3.973$.

Before proceeding, we need to comment on the interpretation of the covariance matrix in the context of universal relations. 
Since the data used for the fitting do not have statistical errors in the usual sense, it is unclear what the physically meaningful uncertainty of the data should be. 
Importantly, one should not confuse statistical errors with the numerical precision of the computed bulk quantities, as the latter goes far beyond what is observationally accessible. 
However, there is always an explicit or implicit choice assigning errors when performing a fit. 
This means that the covariance matrix $\sigma$, in general, does not represent the uncertainty in the universal relation. 
Propagating parameter uncertainties to a universal relation, e.g., sampling them through a multivariate Gaussian centered at the best-fit parameters with a covariance matrix describing their widths, does not reliably estimate the systematic uncertainty. 

In Sec.~S3 in Ref.~\cite{suppmat},
we describe how we introduce an effective calibration parameter (effective error for the fit) such that the new covariance matrix $\hat\sigma$ actually provides an approximation for the systematic error of the universal relation. 
In the following, we will refer to a calibrated (accompanied by the covariance matrix $\hat{\sigma}$) or an uncalibrated (covariance matrix $\sigma$) universal relation.

Last, we comment on the set of bulk quantities present in the proposed universal relation; these are the mass $M$, radius $R$, and the two frequencies $f$ and $p_1$. One might argue that one of the quantities can be eliminated as the three quantities $(M, R, f)$ are universally linked \cite{Andersson:1997rn}. While this is correct to some extent, the spread of the published $f = f(M, R)$ relation is considerably larger than that of our proposed universal relation in Eq.~\eqref{eq:RofMfp1}. This implies that the proposed relation would lose predictional performance if $f = f(M, R)$ is employed to substitute one of the variables. Hence, the proposed relation provides accuracy in addition to already known universal relations.

\section{Application and results}
\label{sec_app_results}

When providing a triple $(M, f, p_1)$ to NPE, one obtains samples from the posterior distribution $p[R|(M, f, p_1)]$ for the radius $R$; we use the mean value of that probability distribution if we are interested in a single value. 
In contrast, a universal relation yields, in general, only one value as a result rather than a posterior distribution or error bars. 
To obtain an error estimate for the prediction of the universal relation, we sample its parameters from a multivariate Gaussian centered at the best-fit values $a_i$ with either covariance matrix $\sigma$ (uncalibrated) or $\hat\sigma$ (calibrated) and then evaluate the ``perturbed'' universal relation Eq.~\eqref{eq:RofMfp1} each time. 
After sampling the parameters 10000 times, we also obtain a ``posterior distribution'' for the radius from the universal relation. 

In the following, we address three key questions: How close are the predicted means to the true values? How wide are the posterior HDIs? And how accurately do the HDIs capture the systematic errors? 

\subsection{Accuracy of predicted radius}
\label{sec_app_11}

\begin{figure}
    \includegraphics[width=1.0\linewidth]{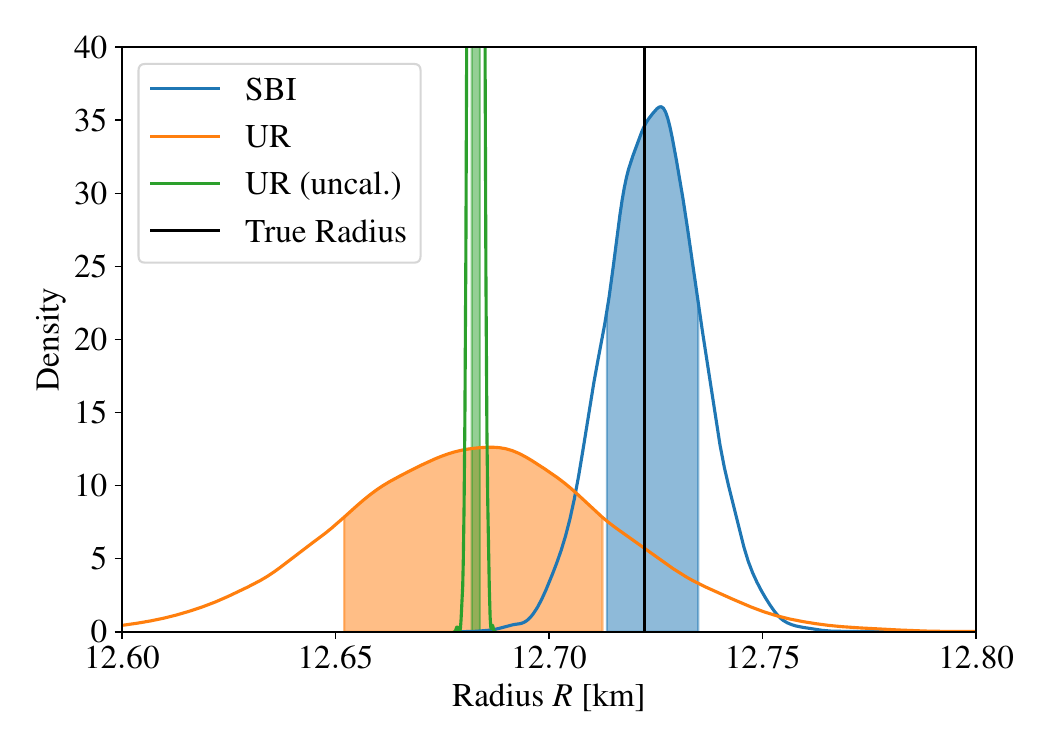}
    \caption{Posterior distributions obtained from applying SBI (blue), calibrated universal relation (orange), and uncalibrated universal relation (green) to the triple of $(M, f, p_1) = (2.33\,M_\odot, 1.87\,\textrm{kHz}, 7.08\,\textrm{kHz})$. 
    The true value of the radius ($12.722\,\textrm{km}$) is shown in black for comparison. 
    The shaded areas represent the 68\,\% HDIs of each distribution.}
    \label{fig:example}
\end{figure}

As a qualitative and straightforward demonstration, we show results for SBI, and both the calibrated and uncalibrated universal relations when applied to a single neutron star that was not used during calibration or fitting (i.e., part of the test data set) in Fig.~\ref{fig:example}. The SBI result includes the true radius well within the 68\% HDI (width of $\approx 22\,\unit{m}$), while the calibrated universal relation covers it at least within the 90\% HDI (the width of the 68\% HDI is $\approx 62\,\unit{m}$). 
The uncalibrated universal relation has by far the smallest HDI (width of $\approx 1.9\,\unit{m}$), however, the true value lies considerably outside the support of the posterior; this demonstrates that the uncalibrated universal relation does not provide reliable information about its systematic error (hence, we will show results only in selected cases), which we will discuss further in Sec.~\ref{sec_app_13}.
The deviation of the mean of SBI from the true radius is only about $2\,\text{m}$, while that of the universal relation is about $39\,\text{m}$.

\begin{figure}
    \includegraphics[width=1.0\linewidth]{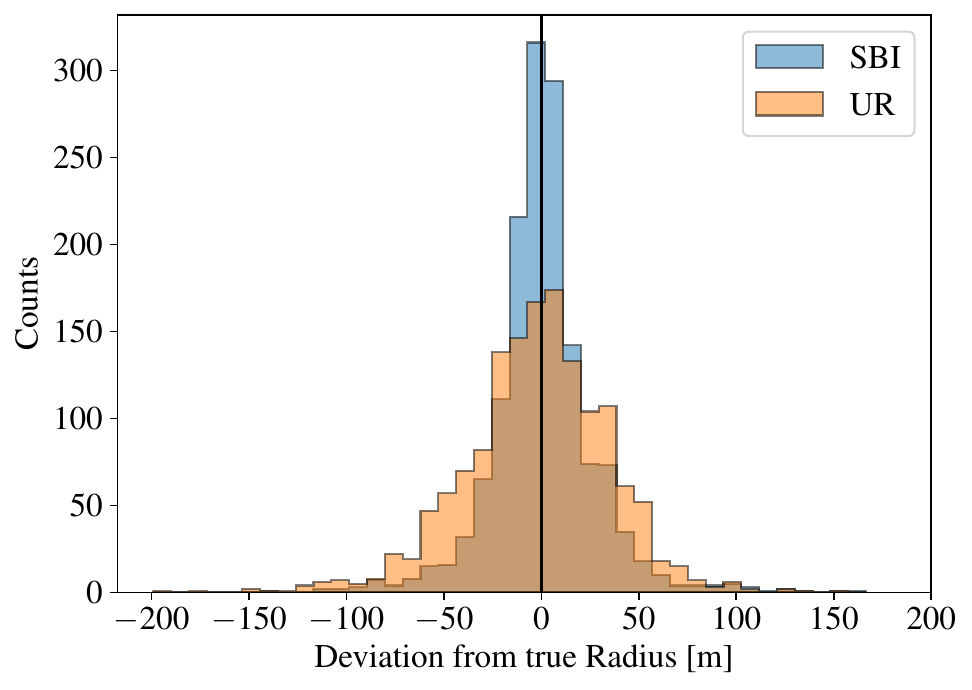}
    \caption{Histograms of the deviations of the mean values of the network's posteriors (SBI) and the calibrated universal relation (UR) from the true value for all data in the test data set.
    }
    \label{fig:hist_deviation}
\end{figure}

To quantitatively assess the closeness of predictions from SBI and the universal relation to the true values, we utilize the entire test dataset (comprising 1470 data points). The deviations of the estimates are shown in Fig.~\ref{fig:hist_deviation}. 
We observe that the deviation is less than $\approx 80\,\unit{m}$ for the majority of data points in both cases, while NPE yields slightly better predictions than the universal relation. 
In particular, the counts of the bins close to vanishing deviation are considerably higher for SBI than for the universal relation.

\subsection{Width of radius HDI}
\label{sec_app_12}

\begin{figure}
    \includegraphics[width=1.0\linewidth]{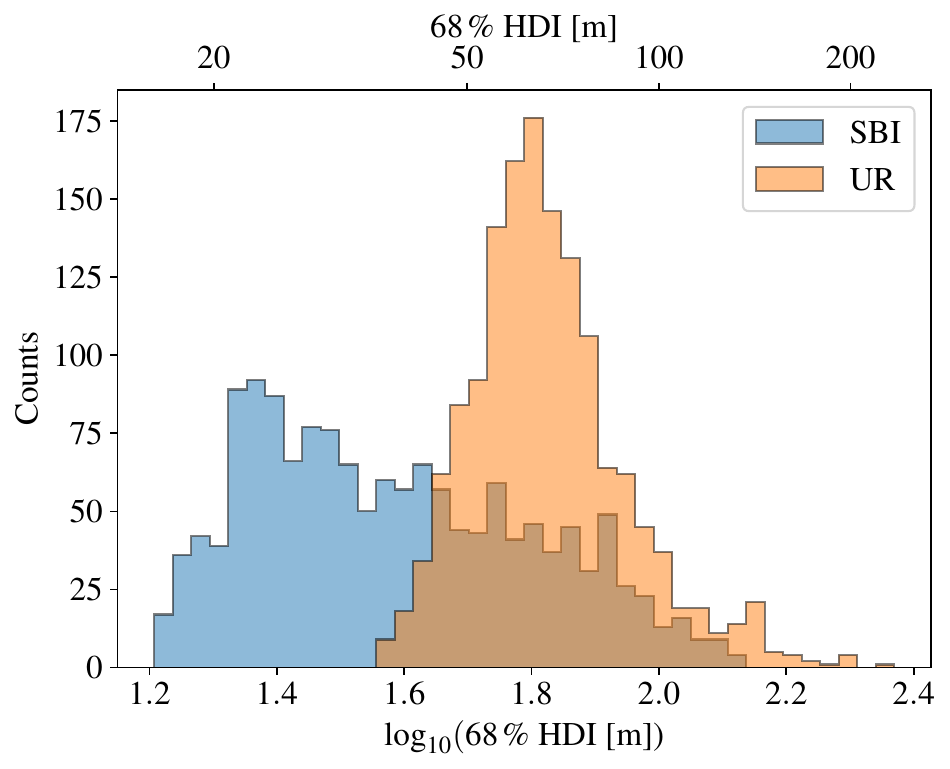}
    \caption{Histograms of the widths of the 68\% 
    HDIs that are returned by the trained network (SBI) and the universal relation (UR) for all data in the test data set. See the main text on how we sample posteriors from the universal relation.}
    \label{fig:hist_hdi_width}
\end{figure}

Another key quantity is the width of the posterior distributions. Ideally, the resulting posteriors are narrow, suggesting small error bars on the estimated radius. We quantify this using the widths of the 68\% HDIs and present a histogram of their values in Fig.~\ref{fig:hist_hdi_width}. We find that the 68\% HDIs for almost all our test data span less than $150\,\unit{m}$.
The posteriors resulting from the universal relation are wider than $40\,\unit{m}$, with most of them having a width of $\approx 70\,\unit{m}$. However, NPE often returns much narrower posteriors that can have a width of as low as $18\,\unit{m}$. 
The uncalibrated universal relation yields considerably narrower posterior widths; however, the previous example already suggests that these are far from reliable. The HDI widths observed in Fig.~\ref{fig:example} reflect the same behavior.

These two checks suggest that SBI provides better estimates (i.e., closer to the true value and with smaller error bars) for the radius than the calibrated universal relation. This is not surprising if we take into account that our neural network optimizes 27510 parameters in the hidden layers, while our universal relation possesses only 5 free parameters. Furthermore, the universal relation is limited by the choice of functional dependence. It will certainly be possible to find a ``better'' universal relation if we allow for more complicated combinations of the variables $M$, $f$, and $p_1$ or consider a wider array of analytic functions such as trigonometric functions or fractional exponents. However, the accuracy of the universal relation is competing with that of the neural network, and it appeals by its high simplicity: The universal relation may be evaluated by hand, while the neural network can be transferred practically only as binary data, and its evaluation requires specific software packages and additional code to be written.

\subsection{Accuracy of systematic error prediction}
\label{sec_app_13}

As a last important question, we now address whether the posterior HDIs predicted by SBI and the universal relation (calibrated and uncalibrated) accurately describe the systematic error.
One necessary condition is that the true values from the test data fall, as often as predicted, into a specified posterior HDI. 
We therefore compute for each element of the test data different HDI intervals and ask whether or not it contains the true value.
The full details of the test are reported in Sec.~S1 in Ref.~\cite{suppmat}.
A subtle complication is that neural network training depends on randomly chosen initial values; consequently, each training (even on the same data) typically yields a different network. In our case, we repeated the training several times, evaluated the calibration diagnostics described in Sec.~S2 in Ref.~\cite{suppmat},
and selected a run with satisfactory calibration. We note that these diagnostics were evaluated on the same test subset used for the final results. This might, in principle, introduce a small network selection bias. 
However, we expect it to be negligible compared to the overall performance difference between the SBI and UR approaches, because the trained networks differ only by random initialization, the calibration metric is inherently noisy, and the dominant uncertainty in the predictive intervals comes from aleatoric uncertainties rather than the network parameters.

\begin{figure}
    \includegraphics[width=1.0\linewidth]{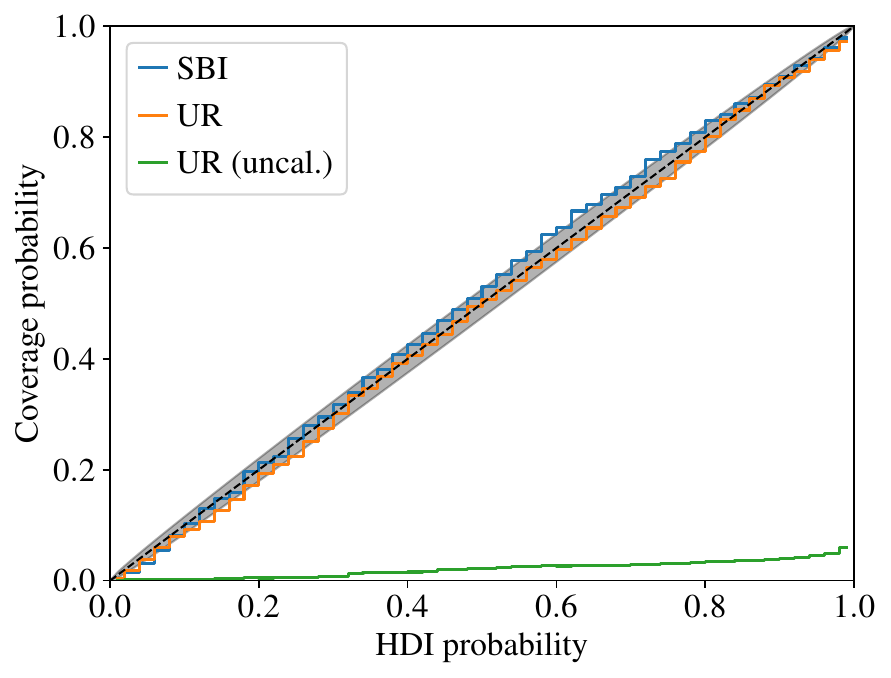}
    \caption{Calibration plot visualizing the reliability of the posteriors of SBI and the universal relation (uncalibrated and calibrated).
        The $x-$axis shows the HDI probability, the $y-$axis shows the coverage probability, i.e., the observed fraction of true values within each HDI. 
        The ideal case corresponds to a diagonal (shown as a black, dashed line) which is very well achieved by the calibrated universal relation; due to the finite accuracy in the SBI network and the finite number of test data, one expects small deviations captured by the binomial proportion confidence interval (shown as blue, solid lines for a 95\,\% confidence level). 
        The uncalibrated universal relation severely underestimates the true posterior uncertainty.
    }
    \label{fig_sbi_test}
\end{figure}

We visualize the results of the test in Fig.~\ref{fig_sbi_test}, where the probability $p$ of the HDI is shown on the $x$-axis, and the fraction of how many times the true value was found within this range, i.e., the coverage, on the $y$-axis. 
The closer the points are to the diagonal line, the more reliable the resulting posteriors are. 
Note that small differences are expected for several reasons, like the finite size of the SBI network and training data, the finite number of posterior samples, and the finite number of test data. 
We indicate binomial uncertainty bands representing an estimate of the expected fluctuations due to the finite number of test data as grey error bars (see Ref.~\cite{suppmat} for details in Section~S1 and SBI calibration in Section~S2). It is also apparent that the uncalibrated universal relation yields hopelessly small error bars. 

\section{Conclusions}

We demonstrated multiple aspects of how SBI can be used as a valuable, simple, and efficient tool for studying neutron stars with universal relations. 
While conventional universal relations provide empirical and accurate estimates for bulk properties of neutron stars in an EOS-insensitive manner, our work is motivated by three practical problems. 
First, finding informative combinations of bulk properties, second, the problem of quantifying the systematic error of such a universal relation, and third, the limitation to a finite set of EOS models when generating neutron star models. 
In this work, we demonstrated that SBI can be used in a systematic, automated way to detect promising parameter combinations, even in a higher-dimensional parameter space. 

The central and novel element of our approach is to utilize SBI to support traditional universal relation construction and use it as a second, independent method. 
SBI only requires one to provide simulated data containing noise, but not to explicitly specify the corresponding likelihood. 
In the context of neutron star universal relations, the data is not equipped with ordinary statistical errors, as they are the result of high-accuracy numerical calculations.
Instead, the effective uncertainties arise due to the imprint of different EOS realizations, which, however, cannot easily be modeled for reasons outlined in Sec.~\ref{methods_ur}. 
In practice, this means that a traditional best-fit approach for finding universal relation fit parameters can provide a point-wise answer, but its covariance matrix does not reflect the systematic error of the universal relation itself. 
Thus, the standard approach to universal relations cannot directly be used to quantify their intrinsic uncertainty (e.g., what are the 68\,\% confidence intervals?).  
However, being able to mitigate the systematic errors is crucial for finding unbiased estimates, which have downstream implications, i.e., finding reliable posteriors for EOS parameters in actual applications. 

In Sec.~\ref{sec_app_results}, we demonstrated that SBI, as implemented in \texttt{sbi}~\cite{tejero-cantero2020sbi, tejero_cantero_2022_8192532, BoeltsDeistler_sbi_2025}, can be used to identify promising bulk property sets by applying it to various combinations of them. 
Using simple summary statistics, one can rank the goodness of correlations and thus decide which combination is worthy of further investigation to construct a universal relation in a traditional way. 
After recovering known universal relations, our method suggested the existence of a new universal relation $R = R(M, f, p_1)$, which we then constructed explicitly. 
We demonstrated that the covariance matrix of such a procedure does, in general, not accurately represent the uncertainties of the universal relation. 
We demonstrate two simple calibration procedures that can be carried out, such that the systematic errors are well represented by a modified covariance matrix or how the posterior can be approximated; see Sec.~S3 in Ref.~\cite{suppmat}
or details. 

Future work should extend our current analysis to include rotating neutron stars. 
There is no technical limitation with respect to the SBI approach; however, the generation of rotating neutron star observables in large quantities can be more challenging, at least for those involving oscillation modes. 
Existing works on the level of universal relations have demonstrated how non-rotating observables can be used to approximate those of rotating ones~\cite{Konstantinou:2022vkr, Kruger:2023olj}. 
In Ref.~\cite{Kruger:2023olj}, systematic errors on the level of EOS parameter estimation due to using slow-rotation approximations, as used in Ref.~\cite{Volkel:2022utc}, have been quantified for different rotation rates. 
Thus, we expect that the SBI predictions are also reliable for slowly rotating neutron stars, at least at few percent level. 
Finally, it would also be interesting to quantify how measurement uncertainties, e.g., those of oscillation modes measured by future detectors such as the Einstein Telescope or Cosmic Explorer, will impact the application of SBI and universal relations.

\acknowledgments
The authors thank Tim Dietrich, Christian Ecker, and Luciano Rezzolla for useful comments on the manuscript. 
S.\,H.\,V. thanks Jonathan Gair and Nihar Gupte for valuable discussions on SBI.
The authors are grateful to the referee for providing insightful comments that helped to improve the manuscript.
S.\,H.\,V. acknowledges funding from the Deutsche Forschungsgemeinschaft (DFG): Project No. 386119226. 
The authors acknowledge support by the High Performance and Cloud Computing Group at the Zentrum für Datenverarbeitung of the University of Tübingen, the state of Baden-W\"urttemberg through bwHPC and the German Research Foundation (DFG) through grant no INST 37/935-1 FUGG.

\bibliography{literature}

@article{Read:2008iy,
    author = "Read, Jocelyn S. and Lackey, Benjamin D. and Owen, Benjamin J. and Friedman, John L.",
    title = "{Constraints on a phenomenologically parameterized neutron-star equation of state}",
    doi = "10.1103/PhysRevD.79.124032",
    journal = "Phys. Rev. D",
    volume = "79",
    pages = "124032",
    year = "2009"
}

@article{tejero-cantero2020sbi,
  doi = {10.21105/joss.02505},
  url = {https://doi.org/10.21105/joss.02505},
  year = {2020},
  publisher = {The Open Journal},
  volume = {5},
  number = {52},
  pages = {2505},
  author = {{\'A}lvaro Tejero-Cantero and Jan Boelts and Michael Deistler and Jan-Matthis Lueckmann and Conor Durkan and Pedro J. Gonçalves and David S. Greenberg and Jakob H. Macke},
  title = {sbi: A toolkit for simulation-based inference},
  journal = {Journal of Open Source Software}
}

@book{sivia2006data,
  title={Data Analysis: A Bayesian Tutorial},
  author={Sivia, D. and Skilling, J.},
  isbn={9780198568315},
  lccn={2006284782},
  series={Oxford science publications},
  url={https://books.google.de/books?id=lYMSDAAAQBAJ},
  year={2006},
  publisher={OUP Oxford}
}

@article{Cook:2006zir,
    author = "Cook, Samantha R. and Gelman, Andrew and Rubin, Donald B.",
    title = "{Validation of Software for Bayesian Models Using Posterior Quantiles}",
    doi = "10.1198/106186006x136976",
    journal = "J. Comp. Graph. Stat.",
    volume = "15",
    number = "3",
    pages = "675--692",
    year = "2006"
}

@misc{talts2020validatingbayesianinferencealgorithms,
      title={Validating Bayesian Inference Algorithms with Simulation-Based Calibration}, 
      author={Sean Talts and Michael Betancourt and Daniel Simpson and Aki Vehtari and Andrew Gelman},
      year={2020},
      eprint={1804.06788},
      archivePrefix={arXiv},
      primaryClass={stat.ME},
      url={https://arxiv.org/abs/1804.06788}, 
}

@misc{tejero_cantero_2022_8192532,
  author       = {Tejero-Cantero, {\'A}lvaro and
                  Boelts, Jan and
                  Deistler, Michael and
                  Lueckmann, Jan-Matthis and
                  Durkan, Conor and
                  Gonçalves, Pedro J. and
                  Greenberg, David S. and
                  Macke, Jakob H.},
  title        = {sbi: A toolkit for simulation-based inference},
  month        = dec,
  year         = 2022,
  publisher    = {Zenodo},
  version      = {v0.21.0},
  doi          = {10.5281/zenodo.8192532},
  url          = {https://doi.org/10.5281/zenodo.8192532},
}

@article{Kashyap:2022wzr,
    author = "Kashyap, Rahul and Dhani, Arnab and Sathyaprakash, Bangalore",
    title = "{Systematic errors due to quasiuniversal relations in binary neutron stars and their correction for unbiased model selection}",
    doi = "10.1103/PhysRevD.106.123001",
    journal = "Phys. Rev. D",
    volume = "106",
    number = "12",
    pages = "123001",
    year = "2022"
}

@misc{Papamakarios:2016ctj,
    author = "Papamakarios, George and Murray, Iain",
    title = "{Fast $\epsilon$-free Inference of Simulation Models with Bayesian Conditional Density Estimation}",
    eprint = "1605.06376",
    archivePrefix = "arXiv",
    primaryClass = "stat.ML",
    month = "5",
    year = "2016"
}

@misc{lueckmann2017flexiblestatisticalinferencemechanistic,
      title={Flexible statistical inference for mechanistic models of neural dynamics}, 
      author={Jan-Matthis Lueckmann and Pedro J. Goncalves and Giacomo Bassetto and Kaan Öcal and Marcel Nonnenmacher and Jakob H. Macke},
      year={2017},
      eprint={1711.01861},
      archivePrefix={arXiv},
      primaryClass={stat.ML},
      url={https://arxiv.org/abs/1711.01861}, 
}

@misc{greenberg2019automaticposteriortransformationlikelihoodfree,
      title={Automatic Posterior Transformation for Likelihood-Free Inference}, 
      author={David S. Greenberg and Marcel Nonnenmacher and Jakob H. Macke},
      year={2019},
      eprint={1905.07488},
      archivePrefix={arXiv},
      primaryClass={cs.LG},
      url={https://arxiv.org/abs/1905.07488}, 
}

@misc{deistler2022truncatedproposalsscalablehasslefree,
      title={Truncated proposals for scalable and hassle-free simulation-based inference}, 
      author={Michael Deistler and Pedro J Goncalves and Jakob H Macke},
      year={2022},
      eprint={2210.04815},
      archivePrefix={arXiv},
      primaryClass={stat.ML},
      url={https://arxiv.org/abs/2210.04815}, 
}

@article{
doi:10.1073/pnas.1912789117,
author = {Kyle Cranmer  and Johann Brehmer  and Gilles Louppe },
title = {The frontier of simulation-based inference},
journal = {Proceedings of the National Academy of Sciences},
volume = {117},
number = {48},
pages = {30055-30062},
year = {2020},
doi = {10.1073/pnas.1912789117},
URL = {https://www.pnas.org/doi/abs/10.1073/pnas.1912789117},
eprint = {https://www.pnas.org/doi/pdf/10.1073/pnas.1912789117}}

@article{Konstantinou:2022vkr,
    author = "Konstantinou, Andreas and Morsink, Sharon M.",
    title = "{Universal Relations for the Increase in the Mass and Radius of a Rotating Neutron Star}",
    doi = "10.3847/1538-4357/ac7b86",
    journal = "Astrophys. J.",
    volume = "934",
    pages = "2",
    month = "6",
    year = "2022"
}

@article{Kruger:2023olj,
    author = {Kr{\"u}ger, Christian J. and V{\"o}lkel, Sebastian H.},
    title = "{Rapidly rotating neutron stars: Universal relations and EOS inference}",
    doi = "10.1103/PhysRevD.108.124056",
    journal = "Phys. Rev. D",
    volume = "108",
    number = "12",
    pages = "124056",
    year = "2023"
}

@article{Volkel:2022utc,
    author = {V{\"o}lkel, Sebastian H. and Kr{\"u}ger, Christian J.},
    title = "{Constraining the nuclear equation of state from rotating neutron stars}",
    doi = "10.1103/PhysRevD.105.124071",
    journal = "Phys. Rev. D",
    volume = "105",
    number = "12",
    pages = "124071",
    year = "2022"
}

@article{OBoyle:2020qvf,
    author = "O'Boyle, Michael F. and Markakis, Charalampos and Stergioulas, Nikolaos and Read, Jocelyn S.",
    title = "{Parametrized equation of state for neutron star matter with continuous sound speed}",
    doi = "10.1103/PhysRevD.102.083027",
    journal = "Phys. Rev. D",
    volume = "102",
    number = "8",
    pages = "083027",
    year = "2020"
}

@article{Greif:2018njt,
    author = "Greif, S. K. and Raaijmakers, G. and Hebeler, K. and Schwenk, A. and Watts, A. L.",
    title = "{Equation of state sensitivities when inferring neutron star and dense matter properties}",
    doi = "10.1093/mnras/stz654",
    journal = "Mon. Not. Roy. Astron. Soc.",
    volume = "485",
    number = "4",
    pages = "5363--5376",
    year = "2019"
}

@article{Annala:2019puf,
    author = {Annala, Eemeli and Gorda, Tyler and Kurkela, Aleksi and N{\"a}ttil{\"a}, Joonas and Vuorinen, Aleksi},
    title = "{Evidence for quark-matter cores in massive neutron stars}",
    reportNumber = "CERN-TH-2019-031, HIP-2019-7/TH",
    doi = "10.1038/s41567-020-0914-9",
    journal = "Nature Phys.",
    volume = "16",
    number = "9",
    pages = "907--910",
    year = "2020"
}

@article{Hinderer:2007mb,
    author = "Hinderer, Tanja",
    title = "{Tidal Love numbers of neutron stars}",
    primaryClass = "astro-ph",
    doi = "10.1086/533487",
    journal = "Astrophys. J.",
    volume = "677",
    pages = "1216--1220",
    year = "2008",
    note = "[Erratum: Astrophys.J. 697, 964 (2009)]"
}

@article{Hinderer:2007mberr,
    author = "Hinderer, Tanja",
    title = "{Erratum: Tidal Love numbers of neutron stars}",
    primaryClass = "astro-ph",
    doi = "10.1088/0004-637X/697/1/964",
    journal = "Astrophys. J.",
    volume = "697",
    pages = "964--965",
    year = "2009",
}

@article{Hartle:1967he,
    author = "Hartle, James B.",
    title = "{Slowly rotating relativistic stars. 1. Equations of structure}",
    doi = "10.1086/149400",
    journal = "Astrophys. J.",
    volume = "150",
    pages = "1005--1029",
    year = "1967"
}

@article{Lindblom:1983ps,
    author = "Lindblom, L and Detweiler, Steven L.",
    title = "{The quadrupole oscillations of neutron stars}",
    doi = "10.1086/190884",
    journal = "Astrophys. J. Suppl.",
    volume = "53",
    pages = "73--92",
    year = "1983"
}

@article{Detweiler:1985zz,
    author = "Detweiler, Steven L. and Lindblom, L.",
    title = "{On the nonradial pulsations of general relativistic stellar models}",
    doi = "10.1086/163127",
    journal = "Astrophys. J.",
    volume = "292",
    pages = "12--15",
    year = "1985"
}

@article{Benhar:1998au,
    author = "Benhar, Omar and Berti, Emanuele and Ferrari, Valeria",
    editor = "Ferrari, V. and Miller, J. C. and Rezzolla, L.",
    title = "{The Imprint of the equation of state on the axial w modes of oscillating neutron stars}",
    doi = "10.1046/j.1365-8711.1999.02983.x",
    journal = "Mon. Not. Roy. Astron. Soc.",
    volume = "310",
    pages = "797--803",
    year = "1999"
}

@article{Andersson:1995wu,
    author = "Andersson, Nils and Kokkotas, Kostas D. and Schutz, Bernard F.",
    title = "{A New numerical approach to the oscillation modes of relativistic stars}",
    doi = "10.1093/mnras/274.4.1039",
    journal = "Mon. Not. Roy. Astron. Soc.",
    volume = "274",
    pages = "1039",
    year = "1995"
}

@article{Yagi:2013bca,
    author = "Yagi, Kent and Yunes, Nicolas",
    title = "{I-Love-Q}",
    eprint = "1302.4499",
    doi = "10.1126/science.1236462",
    journal = "Science",
    volume = "341",
    pages = "365--368",
    year = "2013"
}

@article{Chan:2014kua,
    author = "Chan, T. K. and Sham, Y. -H. and Leung, P. T. and Lin, L. -M.",
    title = "{Multipolar universal relations between f-mode frequency and tidal deformability of compact stars}",
    primaryClass = "gr-qc",
    doi = "10.1103/PhysRevD.90.124023",
    journal = "Phys. Rev. D",
    volume = "90",
    number = "12",
    pages = "124023",
    year = "2014"
}

@article{Lau:2009bu,
    author = "Lau, H. K. and Leung, P. T. and Lin, L. M.",
    title = "{Inferring physical parameters of compact stars from their f-mode gravitational wave signals}",
    eprint = "0911.0131",
    doi = "10.1088/0004-637X/714/2/1234",
    journal = "Astrophys. J.",
    volume = "714",
    pages = "1234--1238",
    year = "2010"
}

@article{Andersson:1997rn,
    author = "Andersson, Nils and Kokkotas, Kostas D.",
    title = "{Towards gravitational wave asteroseismology}",
    reportNumber = "AUTH-GRAV-97-03",
    doi = "10.1046/j.1365-8711.1998.01840.x",
    journal = "Mon. Not. Roy. Astron. Soc.",
    volume = "299",
    pages = "1059--1068",
    year = "1998"
}

@article{Papigkiotis:2023onn,
    author = "Papigkiotis, Grigorios and Pappas, George",
    title = "{Universal relations for rapidly rotating neutron stars using supervised machine-learning techniques}",
    primaryClass = "astro-ph.HE",
    doi = "10.1103/PhysRevD.107.103050",
    journal = "Phys. Rev. D",
    volume = "107",
    number = "10",
    pages = "103050",
    year = "2023"
}

@article{BoeltsDeistler_sbi_2025,
  doi = {10.21105/joss.07754},
  url = {https://doi.org/10.21105/joss.07754},
  year = {2025},
  publisher = {The Open Journal},
  volume = {10},
  number = {108},
  pages = {7754},
  author = {Jan Boelts and Michael Deistler and Manuel Gloeckler and Álvaro Tejero-Cantero and Jan-Matthis Lueckmann and Guy Moss and Peter Steinbach and Thomas Moreau and Fabio Muratore and Julia Linhart and Conor Durkan and Julius Vetter and Benjamin Kurt Miller and Maternus Herold and Abolfazl Ziaeemehr and Matthijs Pals and Theo Gruner and Sebastian Bischoff and Nastya Krouglova and Richard Gao and Janne K. Lappalainen and Bálint Mucsányi and Felix Pei and Auguste Schulz and Zinovia Stefanidi and Pedro Rodrigues and Cornelius Schröder and Faried Abu Zaid and Jonas Beck and Jaivardhan Kapoor and David S. Greenberg and Pedro J. Gonçalves and Jakob H. Macke},
  title = {sbi reloaded: a toolkit for simulation-based inference workflows},
  journal = {Journal of Open Source Software}
}

@article{Musolino:2023edi,
    author = "Musolino, Carlo and Ecker, Christian and Rezzolla, Luciano",
    title = "{On the Maximum Mass and Oblateness of Rotating Neutron Stars with Generic Equations of State}",
    doi = "10.3847/1538-4357/ad1758",
    journal = "Astrophys. J.",
    volume = "962",
    number = "1",
    pages = "61",
    year = "2024"
}

@article{Breu:2016ufb,
    author = "Breu, Cosima and Rezzolla, Luciano",
    title = "{Maximum mass, moment of inertia and compactness of relativistic stars}",
    doi = "10.1093/mnras/stw575",
    journal = "Mon. Not. Roy. Astron. Soc.",
    volume = "459",
    number = "1",
    pages = "646--656",
    year = "2016"
}

@article{Dax:2021tsq,
    author = {Dax, Maximilian and Green, Stephen R. and Gair, Jonathan and Macke, Jakob H. and Buonanno, Alessandra and Sch{\"o}lkopf, Bernhard},
    title = "{Real-Time Gravitational Wave Science with Neural Posterior Estimation}",
    reportNumber = "LIGO-P2100223",
    doi = "10.1103/PhysRevLett.127.241103",
    journal = "Phys. Rev. Lett.",
    volume = "127",
    number = "24",
    pages = "241103",
    year = "2021"
}

@article{Dax:2022pxd,
    author = {Dax, Maximilian and Green, Stephen R. and Gair, Jonathan and P{\"u}rrer, Michael and Wildberger, Jonas and Macke, Jakob H. and Buonanno, Alessandra and Sch{\"o}lkopf, Bernhard},
    title = "{Neural Importance Sampling for Rapid and Reliable Gravitational-Wave Inference}",
    reportNumber = "LIGO-P2200297",
    doi = "10.1103/PhysRevLett.130.171403",
    journal = "Phys. Rev. Lett.",
    volume = "130",
    number = "17",
    pages = "171403",
    year = "2023"
}

@article{Crisostomi:2023tle,
    author = "Crisostomi, Marco and Dey, Kallol and Barausse, Enrico and Trotta, Roberto",
    title = "{Neural posterior estimation with guaranteed exact coverage: The ringdown of GW150914}",
    doi = "10.1103/PhysRevD.108.044029",
    journal = "Phys. Rev. D",
    volume = "108",
    number = "4",
    pages = "044029",
    year = "2023"
}

@article{Dax:2024mcn,
    author = {Dax, Maximilian and Green, Stephen R. and Gair, Jonathan and Gupte, Nihar and P{\"u}rrer, Michael and Raymond, Vivien and Wildberger, Jonas and Macke, Jakob H. and Buonanno, Alessandra and Sch{\"o}lkopf, Bernhard},
    title = "{Real-time inference for binary neutron star mergers using machine learning}",
    reportNumber = "LIGO-P2400294",
    doi = "10.1038/s41586-025-08593-z",
    journal = "Nature",
    volume = "639",
    number = "8053",
    pages = "49--53",
    year = "2025"
}

@article{Rezzolla:2017aly,
    author = "Rezzolla, Luciano and Most, Elias R. and Weih, Lukas R.",
    title = "{Using gravitational-wave observations and quasi-universal relations to constrain the maximum mass of neutron stars}",
    doi = "10.3847/2041-8213/aaa401",
    journal = "Astrophys. J. Lett.",
    volume = "852",
    number = "2",
    pages = "L25",
    year = "2018"
}

@article{Manoharan:2023atz,
    author = "Manoharan, Praveen and Kokkotas, Kostas D.",
    title = "{Finding universal relations using statistical data analysis}",
    doi = "10.1103/PhysRevD.109.103033",
    journal = "Phys. Rev. D",
    volume = "109",
    number = "10",
    pages = "103033",
    year = "2024"
}

@article{Antoniadis:2013pzd,
    author = "Antoniadis, John and others",
    title = "{A Massive Pulsar in a Compact Relativistic Binary}",
    doi = "10.1126/science.1233232",
    journal = "Science",
    volume = "340",
    pages = "6131",
    year = "2013"
}

@article{Bauswein:2017vtn,
    author = "Bauswein, Andreas and Just, Oliver and Janka, Hans-Thomas and Stergioulas, Nikolaos",
    title = "{Neutron-star radius constraints from GW170817 and future detections}",
    doi = "10.3847/2041-8213/aa9994",
    journal = "Astrophys. J. Lett.",
    volume = "850",
    number = "2",
    pages = "L34",
    year = "2017"
}

@article{Raaijmakers:2021uju,
    author = "Raaijmakers, G. and Greif, S. K. and Hebeler, K. and Hinderer, T. and Nissanke, S. and Schwenk, A. and Riley, T. E. and Watts, A. L. and Lattimer, J. M. and Ho, W. C. G.",
    title = "{Constraints on the Dense Matter Equation of State and Neutron Star Properties from NICER{\textquoteright}s Mass{\textendash}Radius Estimate of PSR J0740+6620 and Multimessenger Observations}",
    doi = "10.3847/2041-8213/ac089a",
    journal = "Astrophys. J. Lett.",
    volume = "918",
    number = "2",
    pages = "L29",
    year = "2021"
}

@article{Annala:2017llu,
    author = "Annala, Eemeli and Gorda, Tyler and Kurkela, Aleksi and Vuorinen, Aleksi",
    title = "{Gravitational-wave constraints on the neutron-star-matter Equation of State}",
    reportNumber = "CERN-TH-2017-236",
    doi = "10.1103/PhysRevLett.120.172703",
    journal = "Phys. Rev. Lett.",
    volume = "120",
    number = "17",
    pages = "172703",
    year = "2018"
}

@article{LIGOScientific:2018cki,
    author = "Abbott, B. P. and others",
    collaboration = "LIGO Scientific, Virgo",
    title = "{GW170817: Measurements of neutron star radii and equation of state}",
    reportNumber = "LIGO-P1800115",
    doi = "10.1103/PhysRevLett.121.161101",
    journal = "Phys. Rev. Lett.",
    volume = "121",
    number = "16",
    pages = "161101",
    year = "2018"
}

@article{Kruger:2025sda,
    author = {Kr{\"u}ger, Christian J. and Celato, Mariachiara},
    title = "{Universal relations for fast rotating neutron stars without equation of state bias}",
    eprint = "2509.11882",
    archivePrefix = "arXiv",
    primaryClass = "gr-qc",
    journal = {arXiv e-prints},
    month = "9",
    year = "2025"
}

@book{gsl,
  author    = {M. Galassi and J. Davies and J. Theiler and B. Gough and G. Jungman and M. Booth and F. Rossi},
  title     = {{GNU Scientific Library Reference Manual} (3rd Ed.)},
  year      = {2009},
  publisher = {Network Theory Ltd.},
  isbn      = {0954612078},
  url       = {http://www.gnu.org/software/gsl/},
  note      = {Available at \url{http://www.gnu.org/software/gsl/}},
  publisher = {Network Theory Ltd.},
  address   = {Bristol, UK}
}

@article{Bauswein:2013jpa,
    author = "Bauswein, A. and Baumgarte, T. W. and Janka, H. -T.",
    title = "{Prompt merger collapse and the maximum mass of neutron stars}",
    doi = "10.1103/PhysRevLett.111.131101",
    journal = "Phys. Rev. Lett.",
    volume = "111",
    number = "13",
    pages = "131101",
    year = "2013"
}

@article{Gaertig:2008uz,
    author = "Gaertig, Erich and Kokkotas, Kostas D.",
    title = "{Oscillations of rapidly rotating relativistic stars}",
    doi = "10.1103/PhysRevD.78.064063",
    journal = "Phys. Rev. D",
    volume = "78",
    pages = "064063",
    year = "2008"
}

@article{Kruger:2020gig,
    author = {Kr{\"u}ger, Christian J{\"u}rgen and Foucart, Francois},
    title = "{Estimates for Disk and Ejecta Masses Produced in Compact Binary Mergers}",
    doi = "10.1103/PhysRevD.101.103002",
    journal = "Phys. Rev. D",
    volume = "101",
    number = "10",
    pages = "103002",
    year = "2020"
}

@article{Vretinaris:2019spn,
    author = "Vretinaris, Stamatis and Stergioulas, Nikolaos and Bauswein, Andreas",
    title = "{Empirical relations for gravitational-wave asteroseismology of binary neutron star mergers}",
    doi = "10.1103/PhysRevD.101.084039",
    journal = "Phys. Rev. D",
    volume = "101",
    number = "8",
    pages = "084039",
    year = "2020"
}

@article{Papigkiotis:2025cjy,
    author = "Papigkiotis, Grigorios and Vardakas, Georgios and Likas, Aristidis and Stergioulas, Nikolaos",
    title = "{Universal description of a neutron star{\textquoteright}s surface and its key global properties: A machine learning approach for nonrotating and rapidly rotating stellar models}",
    doi = "10.1103/PhysRevD.111.083056",
    journal = "Phys. Rev. D",
    volume = "111",
    number = "8",
    pages = "083056",
    year = "2025"
}

@article{Papigkiotis:2025dka,
    author = "Papigkiotis, Grigorios and Vardakas, Georgios and Stergioulas, Nikolaos",
    title = "{Assessing Universal Relations for Rapidly Rotating Neutron Stars: Insights from an Interpretable Deep Learning Perspective}",
    eprint = "2508.05850",
    archivePrefix = "arXiv",
    primaryClass = "astro-ph.HE",
    journal = {arXiv e-prints},
    month = "8",
    year = "2025"
}

@article{Fonseca:2021wxt,
    author = "Fonseca, E. and others",
    title = "{Refined Mass and Geometric Measurements of the High-mass PSR J0740+6620}",
    doi = "10.3847/2041-8213/ac03b8",
    journal = "Astrophys. J. Lett.",
    volume = "915",
    number = "1",
    pages = "L12",
    year = "2021"
}

@article{NANOGrav:2019jur,
    author = "Cromartie, H. T. and others",
    collaboration = "NANOGrav",
    title = "{Relativistic Shapiro delay measurements of an extremely massive millisecond pulsar}",
    doi = "10.1038/s41550-019-0880-2",
    journal = "Nature Astron.",
    volume = "4",
    number = "1",
    pages = "72--76",
    year = "2019"
}

@misc{suppmat,
  note = {See Supplemental Material below for additional details supporting the main text. Section S1 describes the posterior coverage analysis, Section S2 presents the methods employed for SBI calibration, and Section S3 details the calibration of universal relations.}
}

\end{document}



\title{Supplemental Material to: \\Combining simulation-based inference and universal relations\\  for precise and accurate neutron star science}

\maketitle

\section{Posterior coverage analysis for SBI and universal relation}
\label{app_accuracy}

To assess the extent to which the SBI posteriors and universal relation capture the systematic uncertainty, we conduct a posterior coverage analysis as follows. 
For each triple of the test data $(M, f, p_1)_k$, we draw 10000 samples from the SBI posterior for $R_k$.
To obtain a corresponding posterior for $R_k$ from the universal relation, we generate 10000 samples of the fit parameters $a_i$ from a multivariate Gaussian distribution centered at the best-fit values and with covariance given by the fit covariance matrix. 
For each sampled parameter set, we evaluate the universal relation to obtain a value of $R_k$.
Equipped with posterior distributions for each of the test data points, we compute the HDI for all integer probabilities $p$ ranging from 0\% to 100\% and check whether the true value $R_k$ falls within each interval. 

For a given probability $p$, this constitutes a Bernoulli trial for each test point; the total number of successes over all $N_{\rm test} = 1470$ data points then follows a binomial distribution with success probability $p$. 
Since $N_{\rm test}$ is finite, we do not expect exactly a fraction $p$ of the points to lie within the respective HDI; instead, we quantify the expected variation using the binomial proportion confidence interval, modeled by
\begin{align}
    \hat{p} = p \pm z_\alpha \sqrt{\frac{p(1-p)}{N_{\rm test}}}\,,
\end{align}
which defines the expected range for the observed fraction of true values falling within the intervals and is shown in Fig.~4 in the main text
as a grey band.
For the 95\,\% confidence level, we have $z_\alpha \approx 1.96$.

\section{SBI calibration using \texttt{sbi} routines}
\label{app:sbc}

The \texttt{sbi} package provides several diagnostic tools to test whether the obtained NPE is well calibrated, and thus can be expected to work reliably. 
We perform some of these tests to complement the checks already carried out in the Section \emph{Application and Results} in the main text.

\begin{figure}[ht]
    \centering
    \includegraphics[width=1\linewidth]{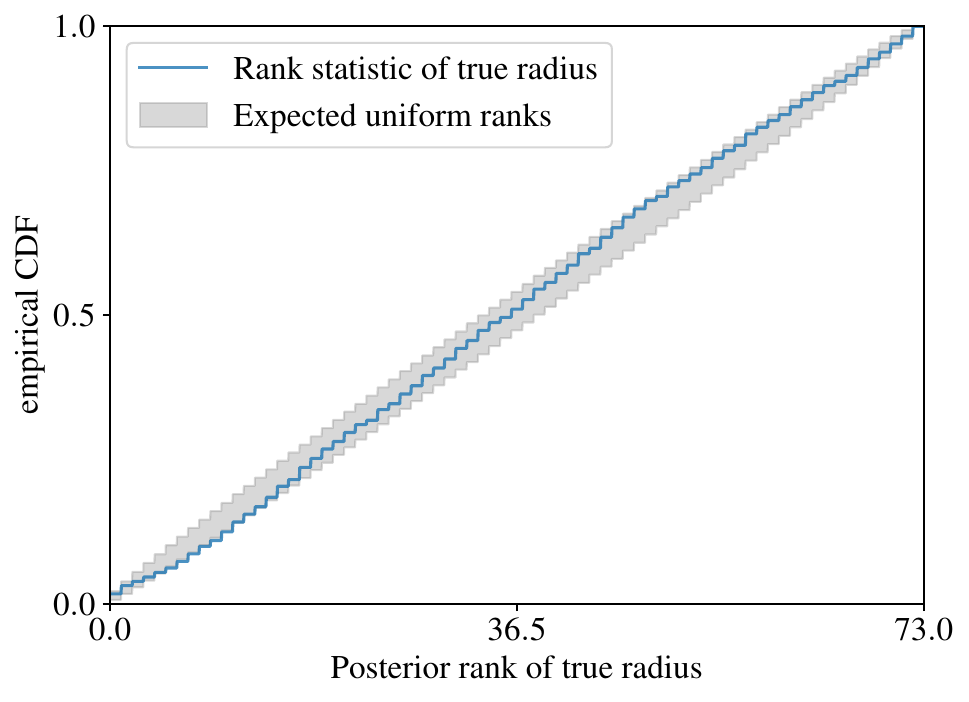}
    \caption{Empirical cumulative density function of the posterior ranks for the true neutron star radius. The grey area shows the 95\% confidence interval for a uniform distribution of ranks.}
    \label{fig:sbc_rankplot}
\end{figure}

\begin{figure}[ht]
    \centering
    \includegraphics[width=1\linewidth]{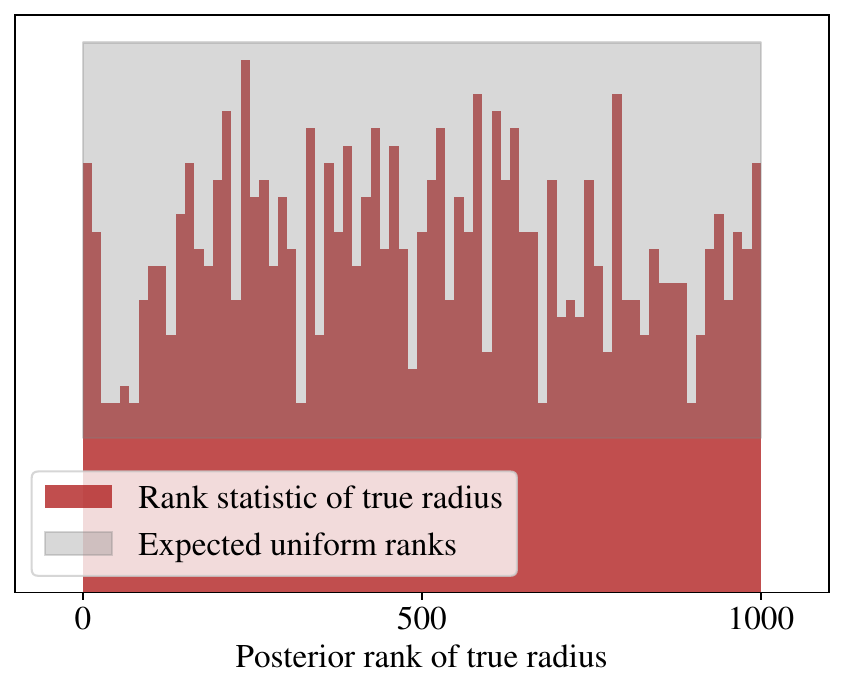}
    \caption{Distribution of the ranks. The grey area denotes the 99\% confidence interval of a uniform distribution.}
    \label{fig:sbc_rank}
\end{figure}

We use the test data and the previously trained NPE for this purpose, and the \texttt{sbi} routines subsequently compute the ranks of the samples, which are employed to perform the tests. 
We show a visualization of the empirical cumulative density function (CDF) in Fig.~\ref{fig:sbc_rankplot} and the distribution of the ranks in Fig.~\ref{fig:sbc_rank}. 
In Fig.~\ref{fig:sbc_rankplot}, the grey area denotes a 95\% confidence interval of a uniform distribution, while in Fig.~\ref{fig:sbc_rank} it denotes a 99\% confidence interval.

Finally, the Kolmogorov-Smirnov test yields a p-value of 13.5\%, indicating that the samples from ranks are indeed drawn from a uniform distribution. 

The built-in tests provided by the \texttt{sbi} package indicate that our neural posterior estimator is well calibrated and will yield reliable results. 
References for the explanation of simulation-based calibration methods used in \texttt{sbi} can be found in Refs.~\cite{Cook:2006zir, talts2020validatingbayesianinferencealgorithms} and references in the \texttt{sbi} package~\cite{tejero-cantero2020sbi, tejero_cantero_2022_8192532, BoeltsDeistler_sbi_2025}.

\section{Calibration of universal relations}
\label{app_ur_calibration}

As noted in Section \emph{Methods}, paragraph \emph{Universal Relation} in the main text,
the covariance matrix obtained by fitting a universal relation to neutron star data that come with EOS noise does not properly reflect its systematic uncertainty. 
However, understanding the latter is crucial for quantifying how reliable the predictions of the universal relation actually are. 

To address this problem, we outline the following strategy, inspired by the SBI calibration test described in Appendix~\ref{app_accuracy}. 
First, we introduce an overall scaling parameter in the least-squares fit, which should reflect the effective error due to EOS noise. Strictly speaking, this procedure inflates the covariance matrix of the fit parameters, which represents epistemic uncertainty, in order to mimic the intrinsic scatter caused by EOS variations, i.e. aleatoric uncertainty. The scaling parameter should therefore be viewed as an empirical calibration device rather than a principled statistical model of the underlying uncertainty.
Since this error cannot be easily obtained from first principles, and would, in general, correlate each data point with each other, we estimate the ``idealized'' error as follows. 
As an approximate measure of how well the universal relation is calibrated, we ask how closely its application in Fig.~4 in the main text
resembles a diagonal. 
By varying the effective error (one single calibration parameter for all data points) and repeating the universal relation fit, we obtain different covariance matrices and different calibration curves.
Finally, we use the value that gives the closest match to the diagonal.

While this procedure gives remarkably good results in our example, there is no guarantee that it also works for more complicated cases.  
One possible interpretation, which requires more quantitative verification and should be considered with caution, is the following. 
The EOS noise in the present case may be sufficiently small that one is effectively in a situation with a large signal-to-noise ratio. 
In standard data analysis applications, posterior distributions at large signal-to-noise ratios are often well described by a Gaussian distribution, which could explain why a single scaling parameter can yield such reliable results in our application. 

\begin{figure}[ht]
    \centering
    \includegraphics[width=1\linewidth]{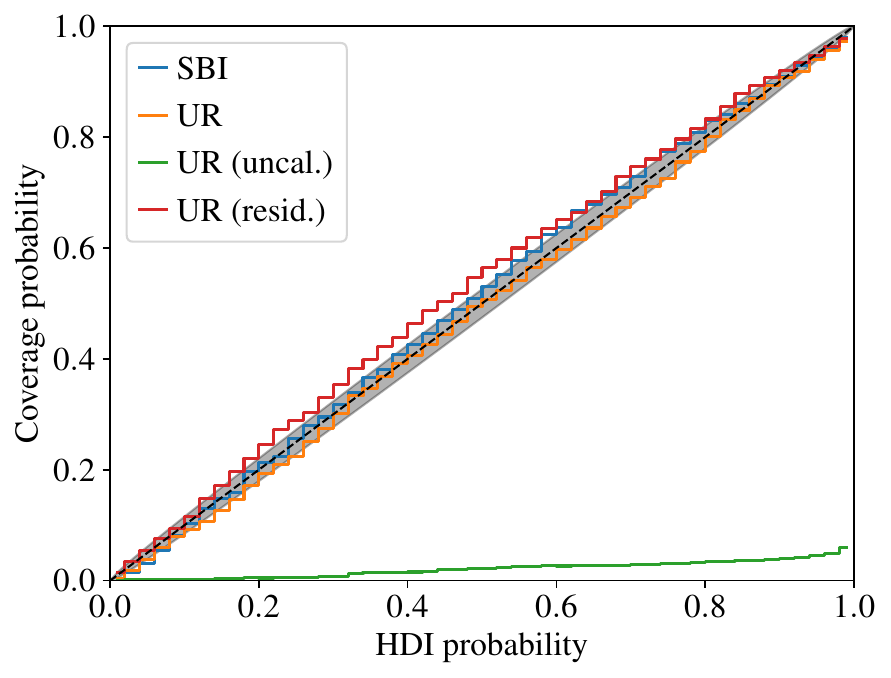}
    \caption{Calibration plot for the universal relation where we assume a Gaussian of constant width (standard deviation of the residuals) for any radius to describe its posterior (red step function). This is a good approximation for the core and tail of the posterior, while it is somewhat too conservative for the central region. We show the calibration plots of the other methods for comparison, too.}
    \label{fig:calibration_std_resid}
\end{figure}

Last, we briefly mention a much simpler approach to estimating the theoretical uncertainty of results obtained from a universal relation. Comparable to the training of a neural network, we split our neutron star data into two sets with sizes of 80\% for fitting and 20\% for testing. After fitting the universal relation to the fitting data set, we determine the residuals for each element of the test data set. We take the standard deviation of the residuals as the width of the Gaussian, centered at the predicted value obtained from the universal relation, to describe the uncertainty of the result that the universal relation yields. Applying this method to the data in this study, we find a standard deviation of the residuals of $\approx 37\,\unit{m}$; assuming that this standard deviation be half the width of the 68\% HDI of the radii, which are obtained by means of the universal relation, results in a surprisingly good calibration. We show the corresponding calibration plot in Fig.~\ref{fig:calibration_std_resid} in red, which contains the calibration plots of the other methods for comparison, too. The calibration of the universal relation using this method is slightly worse compared to the other calibration methods; most notably, for HDI probabilities of roughly 20\% - 70\%, the HDI covers a few more true radii than expected,  meaning that the error bars for this range are somewhat conservative. For smaller as well as larger HDI probabilities, the calibration works well. In summary, this means that our assumption of a Gaussian for the posterior is a good approximation, in particular for its core and tail, while the central region of the posterior should be a bit narrower than the Gaussian. 
We note that this method likely encounter limitations if the actual uncertainties of the radii vary considerably across the data, which is, however, incorporated naturally in SBI.

\bibliography{literature}